\documentstyle[psfig]{mn} 
\begin{document}
\title{Bursting dwarf galaxies from the far-UV and deep surveys}
\author[M. Fioc and B. Rocca-Volmerange]
{Michel~Fioc$^1$\thanks{E-mail: fioc@iap.fr} and Brigitte~Rocca-Volmerange$^{1,2}$\\
$^1$Institut d'Astrophysique de Paris, CNRS,
98 bis Bd. Arago, F-75014 Paris, France \\
$^2$Institut d'Astrophysique Spatiale, B\^at. 121, Universit\'e Paris XI, 
F-91405 
Orsay, France}
\maketitle
\begin{abstract}
The far-ultraviolet (UV) counts and  the deep optical spectroscopic surveys
have revealed an unexpected number of very 
blue galaxies (vBG). Using constraints from the UV and optical, we 
apply the galaxy evolution model PEGASE (Fioc \& Rocca-Volmerange 1997, hereafter FRV) to describe this
population with a cycling star formation. When added to normally evolving galaxy populations, vBG are able to reproduce UV number counts and 
color distributions as well as deep optical redshift distributions fairly 
well. Good
agreement is also obtained with optical counts (including the
Hubble Deep Field). 
The number of vBG is only a small fraction of the
number of normal galaxies, even at faintest magnitudes. In our modelling,
the latter explain the bulk of the excess of faint blue galaxies in an open Universe.
The problem of the blue excess
remains in a flat Universe without cosmological constant.
\end{abstract}
\section{Introduction}
The apparent excess of the number of galaxies at faint magnitudes in the blue relative to predictions of 
non-evolving models, even in the most favourable case of an open Universe, is 
a longstanding 
problem of cosmology.
Various scenarios have been proposed to solve this problem
in a flat Universe, as a strong number density 
evolution of galaxies via merging (Rocca-Volmerange \& Guiderdoni 1990; Broadhurst, Ellis \& Glazebrook 1992) or with a cosmological constant (Fukugita et al. 1990).
In the framework of more conservative pure luminosity evolution models in an 
open Universe, two solutions were
advocated. Either these blue galaxies are intensively star forming galaxies 
at high redshift, or counts are dominated
by a population of intrinsically faint blue nearby galaxies. Looking for the 
optimal luminosity functions (LF) fitting most observational 
constraints, Gronwall \& Koo (1995) have introduced in particular {\em non-evolving populations} of 
faint {\em very} blue galaxies (see also Pozzetti, Bruzual \& Zamorani (1996)),
contributing significantly to faint counts. Such blue colors require
however that {\em individual} galaxies have 
recently been bursting and are thus rapidly evolving. 
With a modelling of the spectral evolution of these galaxies taking also in 
consideration post-burst phases,
Bouwens \& Silk (1996) concluded that the LF adopted by Gronwall \& Koo (1995) leads to a strong excess of nearby galaxies in
the redshift distribution and that vBG may thus not be the main explanation of
the blue excess.

On the basis of considerable observational progress in collecting deep
survey data, it is timely to address the question of the nature of the 
blue excess anew, with the help of our new model PEGASE (FRV).
In this paper, we propose a star formation scenario and a LF
respecting the observational constraints on vBG. 
Far-UV and optical counts are well matched with the classical 
Hubble Sequence population and that bursting population extension.
The importance of vBG relative to normal galaxies and the physical origin of 
bursts are finally discussed in the conclusion.
\section{Observational evidences of very blue galaxies}
In contrast with the so-called `normal' galaxies of the Hubble Sequence,
supposed to form at high redshift with definite star formation timescales,
bursting galaxies are rapidly evolving without clear timescales.
Specifically, in the red post-burst phases, they 
might be undistinguishable from normal
slowly evolving galaxies. The bluest phases during the burst should, however, 
allow to recognize them and to constrain their evolution and their number.

\begin{figure}
\psfig{figure=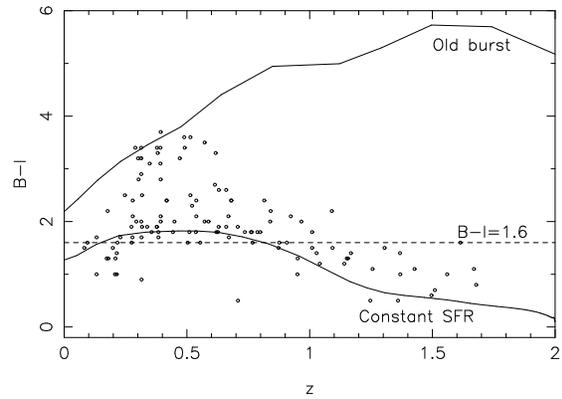,width=8.4cm,angle=-90}
\caption{$B-I$ versus $z$ for galaxies from Cowie et al. (1996) sample.
The thick lines define the envelope of normal galaxies. The upper one holds 
for a 13 Gyr old initial burst without subsequent star formation and the 
lower one for a 10 Gyr old galaxy forming stars at a constant rate. The 
dashed line separates galaxies at $B-I=1.6$. A significant fraction 
of galaxies are observed outside the envelope at $z\sim0.2$, with $B-I<1.6$.}
\label{cowie}
\end{figure}

The existence of galaxies much bluer than normal and classified as starbursts
has been recently noticed
at optical wavelengths by Heyl et al. (1997). At fainter magnitudes ($B=22.5-24$), Cowie et 
al. (1996) deep survey has revealed two populations of blue ($B-I<1.6$) galaxies (Figs.~\ref{cowie} and \ref{nz}). 
Normal star forming galaxies, as predicted by standard models, are observed at high redshift 
($z>0.7$) but another, clearly distinct population of blue galaxies is identified 
at $0<z<0.3$, among which some of them are very blue. 

The best constraint on the weight of these vBG comes from 
the far-UV (2000 \AA) bright counts observed with the balloon experiment 
FOCA2000 (Armand \& Milliard 1994).
By using a standard LF, the authors obtain a strong deficit of predicted galaxies 
in UV counts all along the magnitude range ($UV=14-18$) and argue in favour of a LF biased towards
later-type galaxies. 
With the star formation scenarios and the LF of Marzke et 
al. (1994) fitting optical and near-infrared bright counts (FRV), we confirm 
that this UV deficit reaches a factor 2 (Fig.~\ref{UV}).
Moreover, the $UV-B$ color distributions show
a clear lack of blue galaxies and notably of those with $UV-B<-1.5$ (Fig.~\ref{UV}).
A 10 Gyr old galaxy which formed stars at a constant rate, would 
however only have $UV-B\sim-1.2$. 
Although a low metallicity may lead to bluer colors, it will 
still be too red
and a population of bursting galaxies is clearly needed to explain UV counts 
and the Cowie et al. (1996) data.  
\section{Modeling very blue galaxies}
\subsection{Star formation scenario}
Very blue colors are possible only in very young galaxies or in galaxies currently
undergoing enhanced star formation. Two kinds of models are thus possible and have been advanced by
Bouwens \& Silk (1996) to maintain such a population over a wide range of redshifts. In the first one, 
new blue galaxies are continually formed and leave red fading remnants whereas in the second one,
star formation occurs recurrently. We adopt the last scenario and will discuss
in the conclusion the reasons for this choice. For the sake of simplicity, we assume that 
all vBG form stars periodically. In each period, a burst phase with a constant star formation rate (SFR) 
$\tau_{\rm b}$ and the same initial mass function as in FRV
is followed by a quiescent phase without star formation.
A good agrement with observational constraints is obtained with 100 Myr long burst phases 
occuring every Gyr.
\subsection{Luminosity function}
Because bursting galaxies rapidly redden and fade during inter-burst phases, 
we may not assign a single
LF by absolute magnitude, independently on color. 
We therefore prefer to adopt for vBG a Schechter 
function determined by $\tau_{\rm b}$.

The lack of vBG at $z\ga0.4$ in Cowie et al. (1996) redshift distribution  
is particularly constraining for the LF. It may be interpreted in two 
ways. Either vBG formed only at low redshifts ($z<0.3$) or the lack is due to 
the exponential cut-off in the Schechter LF. 
Physical arguments for such low formation redshifts are weak.
Scenarios invoking a large population of blue dwarf galaxies, as proposed by
Babul \& Rees (1992), generally predict a higher redshift of formation ($z\sim 1$). Adopting the last 
solution, we get 
$M^{\ast}_{\rm b_j}\sim-17$ ($H_0=100\,{\rm km.s^{-1}.Mpc^{-1}}$) for galaxies 
with $B-I<1.6$ and may constrain the 
other parameters of the LF. As noticed by Bouwens \& Silk (1996),
a steep LF extending to very faint magnitudes leads to a large local ($z<0.1$) excess
in the redshift distribution. A steep slope ($\alpha<-1.8$) is however 
only necessary 
to reconcile predicted number counts with observations in a flat universe. In 
an open 
Universe, a shallower slope is possible. In the following, we adopt $\alpha=-1.3$ for vBG.
The normalization is taken in agreement with UV counts and the
Cowie et al. (1996) redshift distribution.
\begin{table}
\begin{tabular}{|c|c|c|c|}
\hline
Galaxy type & $M^{\ast}_{\rm b_j}/\tau^{\ast}_{\rm b}$ & $\alpha$ & $\phi^{\ast}$\\
\hline
E & -20.02 & -1. & $1.91\,10^{-3}$\\ 
S0 & -20.02 & -1. & $1.91\,10^{-3}$\\
Sa & -19.62 & -1. & $2.18\,10^{-3}$\\
Sb & -19.62 & -1. & $2.18\,10^{-3}$\\
Sbc & -19.62 & -1. & $2.18\,10^{-3}$\\
Sc & -18.86 & -1. & $4.82\,10^{-3}$\\
Sdm & -18.86 & -1. & $9.65\,10^{-3}$\\
vBG & $3.95\,10^5$ & -1.3 & $6.63\,10^{-2}$\\
\hline
\end{tabular}
\caption{Luminosity functions parameters ($H_0=100\,{\rm km.s^{-1}.Mpc^{-1}}$). For vBG,
we give the SFR during the burst phase $\tau_{\rm b}^{\ast}$ at the LF knee in $M_{\odot}.{\rm Myr}^{-1}$.}
\label{FL}
\end{table}
\begin{figure}
\psfig{figure=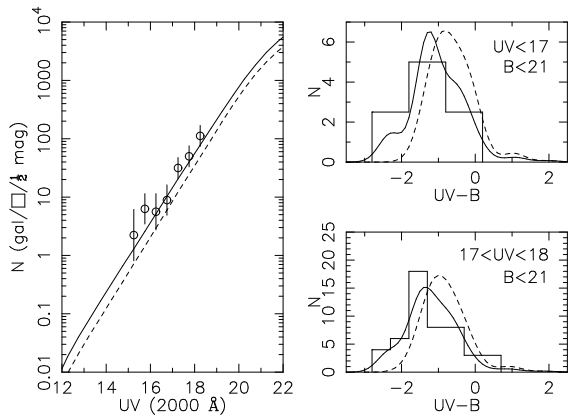,width=8.4cm,angle=-90}
\caption{Number counts and color distributions predicted with Marzke et al. (1994) LF
(dashed) and with vBG (see text) and observed Heyl et al. (1997) LFs (solid)
compared to the observations
of Armand \& Milliard (1994) (circles and histograms). Color distributions are 
normalized to the areas of the histograms.} 
\label{UV}
\end{figure}
\begin{figure}
\psfig{figure=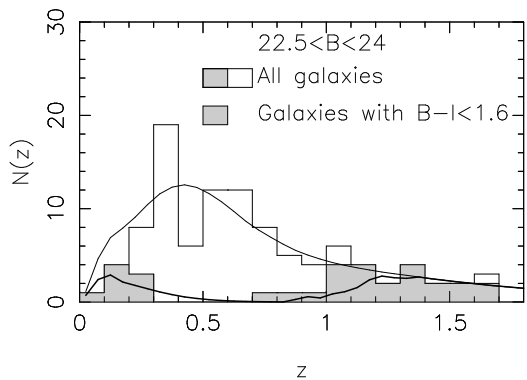,width=8.4cm,angle=-90}
\caption{Predicted redshift distribution ($22.5<B<24$) compared to the observations
of Cowie et al. (1996). The thick line is for galaxies with $B-I<1.6$ and
the thin line for all galaxies.}
\label{nz}
\end{figure}
\section{Galaxy counts}
\begin{figure}
\psfig{figure=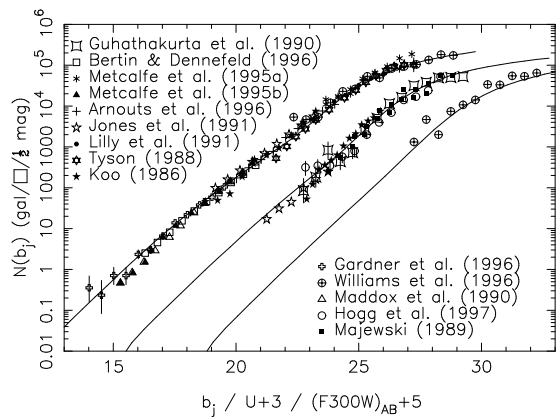,width=8.4cm,angle=-90}
\caption{Number counts in $b_{\rm j}$ (left), $U$ (middle) and $F300W$ (right).}
\label{compt}
\end{figure}

\begin{figure}
\psfig{figure=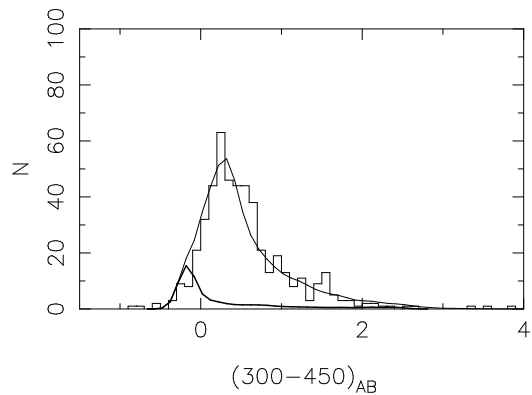,width=8.4cm,angle=-90}
\caption{$F300W-F450W$ color distribution in the HDF for 
$(F300W)_{\rm AB}<27.75$ and $(F450W)_{\rm AB}<28.75$ at 80 per cent completeness. Thin line: all galaxies.
Thick line: vBG only.}
\label{HDF}
\end{figure}

\begin{figure}
\psfig{figure=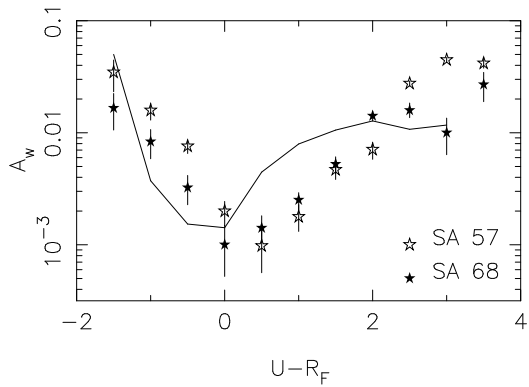,width=8.4cm,angle=-90}
\caption{Amplitude at $1\degr$ of the angular correlation function in $B_{\rm J}=20-23.5$
as a function of $U-R_{\rm F}$ color in bins of 1 magnitude wide. Stars are 
from Landy et al. (1996). The solid line is the amplitude predicted
without evolution of the intrinsic clustering in proper coordinates.
}
\label{Aw}
\end{figure}

Galaxy counts and the amplitude of the projected correlation function by color
in an open Universe ($\Omega_0=0.1$, $\lambda_0=0$, $H_0=65\,{\rm km.s^{-1}.Mpc^{-1}}$), obtained with our  
modelling of vBG and the standard scenarios\footnote{A constant SFR and $z_{\rm for}=2$
are assumed for Sd-Im galaxies.} discussed in FRV, are presented in Fig.~\ref{UV} to \ref{Aw}.
For `normal' types, we use the $z=0$ SSWML LF of Heyl et al. (1997), after 
deduction of the contribution of vBG. Characteristics of the LF finally adopted 
are given in table~\ref{FL}.
Though faint in the blue, vBG play an essential role in UV bright counts thanks to their 
blue $UV-B$ colors and give a much better agreement on 
Fig.~\ref{UV}, both in number counts and color distributions. 

Their contributions to counts at longer wavelengths is however
much smaller. They represent less than 10 per cent of the total number of galaxies at $B=22.5-24$ 
in Cowie et al. (1996) redshift survey and may thus not be the main explanation
of the excess of faint blue galaxies observed over the model without evolution. High redshift,
intrinsically bright galaxies forming stars at a higher rate in the past are the main reason
as it clearly arises from the $z>1$ tail of normally blue galaxies.
In an open Universe, these galaxies reproduce the faint $B$ and even $U$ 
counts, assuming a normalization
of the LF fitting the bright counts of Gardner (1996) as discussed in FRV.
The agreement with the Hubble Deep Field (HDF, Williams et al. 1996) in the blue is notably 
satisfying. Though a small deficit may be observed in the $F300W$ band (3000 \AA),
the $F300W-F450W$ (3000\AA--4500\AA) color distribution is well reproduced (Fig.~\ref{HDF}). The fraction of vBG at these faint magnitudes
is still small; they are therefore not the main reason for the agreement 
with HDF data.

From this previous analysis, it is clear that vBG are difficult to constrain in the visible
from broad statistics like number counts and even color distributions.
The angular correlation function might be promising since it is more directly
related 
to the redshift distribution. In a $B_{\rm J}=20-23.5$ sample, 
Landy, Szalay \& Koo (1996) recently obtained an 
unexpected increase of the amplitude $A_w$
of the angular correlation function with galaxy colors 
$U-R_{\rm F}<-0.5$, and suggested that this might be due to a population of 
vBG located at $z<0.4$. 
We compute $A_w$ from our redshift distributions, assuming the classical
power law
for the local spatial correlation function and no evolution of the intrinsic 
clustering in proper coordinates.
A slope $\gamma=1.8$ and a single correlation length 
$r_0=5.4h^{-1}\,{\rm Mpc}$ (see Peebles (1993)) are adopted for all types. 
The increase of $A_w$ in the blue 
naturally arises from our computations (Fig.~\ref{Aw}) and is due to vBG.
The interval of magnitude, the faint $M^{\ast}$ and the color criterion conspire to select
galaxies in a small range of redshift.
In spite of the simplicity of our computation of $A_w$, 
the trend we obtain is very 
satisfying. 
Modelling improved by extra physics or type effects might better fit the $A_w$-color
relation, but at
the price of 
an increased number of parameters. 
\section{Conclusion}
We modelled the vBG appearing
notably in UV counts with cycling star formation. 
Our modelling agrees
well with the constraints brought by the 2000\AA\ bright counts 
(Armand \& Milliard 1994), 
the redshift survey of Cowie
et al. (1996) and the angular correlation function of Landy et al. (1996).
The cycling star formation provides very blue colors
in a more physical way than by assuming a population of unevolving galaxies. 
The continual formation of new bursting galaxies
might lead to similar predictions in the UV-optical, but would produce a high
number of very faint red remnants. Future deep near-infrared surveys 
should provide discriminations between these scenarios.  
The hypothesis of cycling star forming galaxies has however some theoretical
support. 
The feedback of supernovae on the interstellar medium,
may lead to 
oscillations of the SFR (Wiklind 1987; Firmani \& Tutukov 
1993; Li \& Ikeuchi 1988). Since the probability of propagation 
of star formation increases with galaxy mass (Coziol 1996),
according to the stochastic self propagation star formation theory 
(Gerola, Seiden \& Schulman 1980),
this behaviour should be more frequent in small 
galaxies. 
More regular SFR might be attained in 
more massive ones.
The nature
of vBG is poorly constrained, but we tentatively identify them 
from their typical luminosity and ${\rm H}\alpha$ equivalent width 
($\sim 200$ \AA) with H{\sc ii} galaxies (Coziol 1996).

Very blue galaxies, as modelled in this paper, are only a small 
fraction of the number of
galaxies predicted at faint magnitudes in the visible and are not 
the main reason
for the excess of blue galaxies, although they may cause some confusion
in the interpretation of the faint surveys. In an open Universe, the population
of normal high redshift star forming galaxies, 
even with a nearly flat LF, reproduces fairly well the counts till the faintest magnitudes
observed by the Hubble Space Telescope. 
As is now well established, this population is however, unable to explain the excess
of faint blue galaxies in a flat Universe. Increasing strongly
the number of vBG (for example, with a steeper slope of the LF) may not be the solution
since it would lead to an excess of galaxies at very low redshift which is not observed. 
This result depends however on the hypotheses of pure luminosity evolution
and null cosmological constant. A flat Universe might still be possible
if other evolutionary scenarios are favoured by new observations in the 
far-infrared and submillimeter.


\begin{thebibliography}{}
\bibitem[]{}
Armand C., Milliard B., 1994, A\&A 282, 1
\bibitem[]{}
Arnouts S., de Lapparent V., Mathez G., Mazure A., Mellier Y., Bertin E., 
Kruszewski A., 1996, A\&AS (in press)
\bibitem[]{}
Babul A., Rees M. J., 1992, MNRAS 255, 346
\bibitem[]{}
Bertin E., Dennefeld M., 1997, A\&A 317, 43
\bibitem[]{}
Bouwens R. J., Silk J., 1996, ApJ 471, L19
\bibitem[]{}
Broadhurst T. J., Ellis R. S., Glazebrook K., 1992, Nature~355,~55
\bibitem[]{}
Coziol R., 1996, A\&A 309, 345
\bibitem[]{}
Cowie L. L., Songaila A., Hu E. M., Cohen J. G., 1996, AJ~112,~839
\bibitem[]{}
Fioc M., Rocca-Volmerange B., 1997 (accepted)
\bibitem[]{}
Firmani C., Tutukov A. V., 1994, A\&A 288, 713
\bibitem[]{}
Fukugita M., Yamashita K., Takahara F., Yoshii Y., 1990, ApJ 361, L1
\bibitem[]{}
Gardner J. P., Sharples R. M., Carrasco B. E., Frenk C. S., 1996, MNRAS 282, L1
\bibitem[]{}
Gerola H., Seiden P. E., Schulman L. S., 1980, ApJ 242, 517
\bibitem[]{}
Gronwall C., Koo D. C., 1995, ApJ 440, L1
\bibitem[]{}
Guhathakurta P., Tyson J. A., Majewski S. R., 1990, in
Kron R. G., ed, Astronomical Society of the Pacific, San Francisco, 
Evolution of the 
Universe of Galaxies, p. 304
\bibitem[]{}
Heyl J., Colless M., Ellis R. S., Broadhurst T., 1997, MNRAS 285, 613
\bibitem[]{}
Hogg D. W., Pahre M. A., McCarthy J. K., Cohen J. G., Blandford R., Smail I., 
Soifer B. T., 1997 (astro-ph/9702241)
\bibitem[]{}
Jones L. R., Fong R., Shanks T., Ellis R. S., Peterson, B. A., 1991, MNRAS 249, 481
\bibitem[]{}
Koo D. C., 1986, ApJ 311, 651
\bibitem[]{}
Landy S. D., Szalay A. S., Koo D. C., 1996, ApJ 460, 94
\bibitem[]{}
Li F., Ikeuchi S., 1989, PASJ 41, 221
\bibitem[]{}
Lilly S. J., Cowie L. L., Gardner J. P., 1991, ApJ 369, 79
\bibitem[]{}
Maddox S. J., Sutherland W. J., Efstathiou G., Loveday J., Peterson B. A., 1990, 
MNRAS 241, 1p
\bibitem[]{}
Majewski S. R., 1989, in Frenk C.S. et al., eds, Proc. NATO,
The Epoch of Galaxy Formation, Dordrecht, Kluwer, p. 86
\bibitem[]{}
Marzke R. O., Geller M. J., Huchra J. P., Corwin H. G., 1994, AJ 108, 437
\bibitem[]{}
Metcalfe N., Shanks T., Fong, R., Roche N., 1995a, MNRAS~273,~357
\bibitem[]{}
Metcalfe N., Fong R., Shanks T., 1995b, MNRAS 274, 769
\bibitem[]{}
Peebles P. J. E., 1993, Principles of Physical Cosmology, Princeton 
University Press, Princeton
\bibitem[]{}
Pozzetti L., Bruzual A. G., Zamorani G., 1996, MNRAS 281, 953
\bibitem[]{}
Rocca-Volmerange B., Guiderdoni B., 1990, MNRAS 247, 166
\bibitem[]{}
Tyson J. A., 1988, AJ 96, 1
\bibitem[]{}
Wiklind T., 1987, in Thuan T. X., Montmerle T., Tran Thanh Van J., eds,
Starbursts and Galaxy Evolution, Fronti\`eres, Gif-sur-Yvette, p. 495
\bibitem[]{}
Williams R. E. et al., 1996, AJ 112, 1335
\end{thebibliography}
\end{document}